\documentclass{ws-procs9x6}

\begin{document}

\title{The Clustering of XMM-{\em Newton} Hard X-ray Sources}

\author{M. Plionis$^{1,2}$ \footnote{\uppercase{W}ork partially
supported by grant \uppercase{CONAC}y\uppercase{T}-2002-C01-39679 
of the \uppercase{M}exican Government} \;
S.Basilakos$^1$, A.Georgakakis$^1$, I.Georgantopoulos$^1$}

\address{$^1$ Institute of Astronomy \& Astrophysics - 
National Observatory of Athens, Greece \\
$^2$ Instituto Nacional de Astrof\'{\i}sica, \'Optica y Electr\'onica, 
Puebla, M\'exico}


\maketitle

\abstracts{
We present the clustering properties of hard (2-8\,keV) X-ray
selected sources detected in a wide field ($\approx \rm 2\,deg^{2}$)
shallow [$f_X(\rm 2-8\,keV)\approx 10^{-14}\rm \, erg \, cm^{-2} \,
s^{-1}$] and contiguous  XMM-{\it Newton} survey. We perform an angular
correlation function analysis using a total of  171 sources to the
above flux limit.  We detect a $\sim 4\sigma$ correlation signal out to
300\,arcsec with $w(\theta < 300^{''})\simeq  0.13 \pm 0.03$. Modeling
the two point correlation function as a  power law of the form 
$w(\theta)=(\theta_{\circ}/\theta)^{\gamma-1}$ we find:
$\theta_{\circ}=48.9^{+15.8}_{-24.5}$ arcsec and $\gamma=2.2\pm
{0.30}$. Fixing the correlation function slope to $\gamma=1.8$ we
obtain  $\theta_{\circ}=22.2^{+9.4}_{-8.6}$\,arcsec. Using 
Limber's integral equation and a variety of possible luminosity functions of
the hard X-ray population, we find a relatively large correlation
length, ranging from $r_{\circ}\sim 9$ to 19 $h^{-1}$ Mpc (for
$\gamma=1.8$ and the {\em concordance} cosmological model), with this range
reflecting also different evolutionary models for the source
luminosities and clustering characteristics.}

\section{Introduction}
The overall knowledge of the AGN clustering using X-ray data 
comes mostly from the soft X-ray band 
(Boyle \& Mo 1993; Vikhlinin \& Forman 1995; Carrera et al. 1998; 
Akylas, Georgantopoulos, Plionis, 2000; Mullis 2002), which is however biased
against absorbed AGNs. 
Recently, Yang et al. (2003)   
performing a counts-in cells analysis of a
deep ($f_{2-8 keV} \sim 3 \times 10^{-15}$ erg s$^{-1}$ cm$^{-2}$)  
{\em Chandra} survey in the Lockman Hole North-West region,
found that the hard band sources are highly clustered 
with $\sim$ 60$\%$ of them being distributed in overdense regions.

In this study we use a hard (2-8\,keV) X-ray selected sample, compiled
from a shallow (2-10\,ksec per pointing) XMM-{\em Newton} survey 
near the North and South Galactic Pole regions. A total of 18 pointings were
observed out of which only 5 were discarded due to elevated particle
background at the time of  the observation. 
A full description of the data  reduction, source
detection and flux estimation are presented by Georgakakis et
al. (2003). Here we just note that it comprises of 171 sources above the
$5\sigma$ detection threshold to the limiting flux of $f_X(\rm
2-8\,keV) \approx 10^{-14}\, erg \, s^{-1}\, cm^{-2}$. 

\section{Correlation function analysis}
In the present study we use as the estimator of the 2-point
correlation function the following:
$w(\theta)=f(N_{DD}/N_{DR})-1$, 
where $N_{DD}$ and $N_{DR}$ is the number of data-data and data-random
pairs respectively at separations $\theta$ and  $\theta +
d\theta$. In the above relation $f$ is the normalization factor
$f = 2 N_R /(N_D-1)$ where $N_D$ and $N_R$ are the total number of
data and random points respectively. 
To account for the different source selection
and edge effects, we have produced 100 Monte Carlo random realizations of the 
source distribution within the area of the survey by taking into
account variations in sensitivity which might affect the correlation
function estimate. Indeed, the flux threshold for detection depends on
the off-axis angle from the center of each of the XMM-{\it Newton} 
pointings. Since
the random catalogues must have the same selection effects as the real 
catalogue, sensitivity maps are used to discard random points in
less sensitive areas (close to the edge of the pointings). 
This is accomplished, to  the first approximation, by
assigning a flux to the random points using the Baldi et al. (2002)
2-10\,keV $\log N - \log S$ (after transforming to the 2-8\,keV band
assuming $\Gamma=1.7$). If the  flux of a random point is less than
5 times the local {\em rms}  noise (assuming Poisson statistics for the
background) the point is excluded from the random data-set. We note
that the Baldi et al. (2002) $\log N - \log S$ is in good agreement
with the 2-8\,keV number counts estimated in the present survey. 
\begin{figure}[ht]
\centerline{\epsfxsize=2.5in \epsfbox{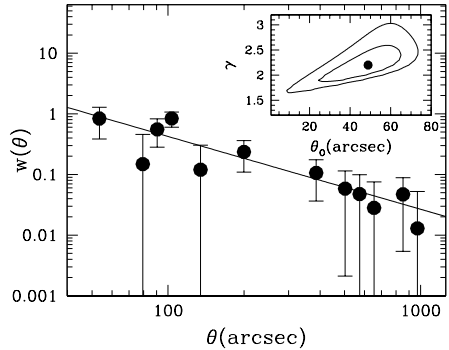}}   
\caption{The 2-point angular correlation function 
of the hard (2-8 keV) X-ray sources. 
Insert: Iso-$\Delta \chi^{2}$ contours in the $\gamma$-$\theta_{\circ}$ 
parameter space.}
\end{figure}

The results of our analysis are 
shown in Figure 1, were the line corresponds to the best-fit power law model 
$w(\theta)=(\theta_{\circ}/\theta)^{\gamma-1}$ using the standard
$\chi^{2}$ minimization procedure in which each correlation point is
weighted by its error.  We find a statistically significant signal 
with $w(\theta < 300^{''})\simeq  0.13 \pm 0.03$
at the 4.3$\sigma$  confidence level using Poissonian errors.
The best fit clustering parameters are:
$\theta_{\circ}=48.9^{+15.8}_{-24.5}$ and $\gamma=2.2\pm
{0.30}$, where the errors correspond to $1\sigma$ 
uncertainties. Fixing the correlation function slope to its
nominal value, $\gamma=1.8$, we estimate
$\theta_{\circ}=22.2^{+9.4}_{-8.6}$ arcsec.

Our results show that hard X-ray sources are strongly 
clustered, even more  than the soft ones (see Vikhlinin 
\& Forman 1995; Yang et al. 2003; Basilakos et al. in preparation). 
Our derived angular correlation length $\theta_{\circ}$ is in rough agreement,
although somewhat smaller (within 1$\sigma$) with the
{\em Chandra} result of $\theta_{\circ}=40 \pm 11$ arcsec
(Yang et al. 2003). 
The stronger angular clustering with respect to the soft
sources could be either due to the higher flux limit of the hard
XMM-{\em Newton} sample, resulting in the selection of relatively 
nearby sources, or could imply an association of our hard X-ray
sources with high-density peaks

\section{The spatial correlation length using $w(\theta)$}
The angular correlation function $w(\theta)$ can be obtained
from the spatial one, $\xi(r)$, through the Limber transformation
(Peebles 1980). If the spatial correlation function is
modeled as
$\xi(r,z)=(r/r_{\circ})^{-\gamma} (1+z)^{-(3+\epsilon)}$.
The angular amplitude $\theta_{\circ}$ is related to the 
correlation length $r_{\circ}$ in 
three dimensions via Limber's equation.
Note that if $\epsilon=\gamma-3$, the clustering is constant in comoving 
coordinates (comoving clustering) while if $\epsilon=-3$ the 
clustering is constant in physical coordinates.
We perform the Limber's  inversion in the framework of the {\em
  concordance} $\Lambda$CDM cosmological model 
($\Omega_{\rm m}=1-\Omega_{\Lambda}=0.3$, $H_{\circ}=70$km s$^{-1}$ 
Mpc$^{-1}$).

The expected redshift distribution and the predicted 
total number, $N$, of the X-ray sources which enters in Limber's 
integral equation can be found using the hard band luminosity
functions of Ueda et al. (2003).
We also use different models for the evolution of the hard
X-ray sources: a pure luminosity evolution (PLE) or the more realistic
luminosity dependent density evolution (LDDE; Ueda et al 2003). 
The LDDE model with respect to the PLE gives an expected redshift 
distribution shifted to larger redshifts, with a 
median redshift of $\bar{z} \simeq 0.75$. 

For the comoving clustering model ($\epsilon=\gamma-3$) 
and using the LDDE evolution model, we estimate the hard X-ray source
correlation length to be: $r_{\circ}=19 \pm 3 \; h^{-1}$ Mpc
and $r_{\circ}=13.5 \pm 3 \; h^{-1}$ Mpc 
for $\gamma=1.8$ and $\gamma=2.2$ respectively.
While if $\epsilon=-3$ the corresponding values
are: $r_{\circ}=11.5 \pm 2 \; h^{-1}$ Mpc and 
$r_{\circ}=6 \pm 1.5 \; h^{-1}$ Mpc, respectively.

The estimated clustering lengths (for $\gamma=1.8$)
are a factor of $\gtrsim 2$ larger than the 
corresponding values of the {\em 2QZ} QSO's (Croom et al. 2001). However, 
the most luminous, and thus nearer, {\em 2QZ} sub-sample ($18.25<b_{j}<19.80$) 
has a larger correlation length ($\sim 8.5 \pm 1.7 \; h^{-1}$ Mpc) than the
overall sample (Croom et al. 2002), in marginal agreement 
with our $\epsilon=-3$ clustering evolution results.

The large spatial clustering length of our hard X-ray sources 
can be compared with that of Extremely Red Objects and luminous
radio  sources (Roche, Dunlop \& Almaini 2003; Overzier et 
al. 2003; R\"{o}ttgering et al. 2003) which are found to be 
in the range $r_{\circ} \simeq 12 - 15 \; h^{-1}$ Mpc. 


\end{document}